\documentclass[11pt]{article} 
\usepackage[a4paper, margin=1.25in]{geometry} 
\usepackage[T1]{fontenc}

\usepackage[usenames,dvipsnames,svgnames,x11names]{xcolor}
\usepackage[nocompress]{cite}

\usepackage[pdftex]{graphicx}
\graphicspath{{./figures/}}
\DeclareGraphicsExtensions{.pdf}

\usepackage[cmex10]{amsmath}
\usepackage{amssymb}
\usepackage{mathtools}
\usepackage{amsfonts}
\usepackage{gensymb}

\usepackage[caption=false]{subfig}
\usepackage{multirow}
\usepackage{booktabs}
\usepackage{colortbl}
\usepackage{tablefootnote}

\usepackage{array}
\newcolumntype{L}[1]{>{\raggedright\let\newline\\\arraybackslash\hspace{0pt}}m{#1}}
\newcolumntype{R}[1]{>{\raggedleft\let\newline\\\arraybackslash\hspace{0pt}}m{#1}}

\usepackage[pdftex,colorlinks=true,urlcolor=blue,citecolor=blue]{hyperref}

\usepackage{enumitem}

\begin{document}

\title{\huge Per-Phase Fidelity Attribution for Quantum Compilers using HBR Decomposition}
\author{
Chandrachud Pati and Yogesh Simmhan
\\~\\
\textit{Indian Institute of Science, Bangalore 560012 India}
\\~\\
\texttt{Email: \{chandrachudp, simmhan\}@iisc.ac.in}
}
\date{}

\maketitle

\begin{abstract}
Quantum compilers sit between an algorithm's theoretical promise and what executes on physical hardware. 
Existing benchmarks report aggregate post-transpilation metrics but cannot attribute where fidelity is lost within the compilation pipeline. 
We present \emph{HBR decomposition}, a per-phase fidelity attribution model that quantifies relative fidelity loss across \underline{H}igh-level structural decomposition (H), \underline{B}asis translation (B), and \underline{R}outing (R).
We evaluate three production SDKs (Qiskit, PennyLane, TKET) across eight algorithms on two backend topologies: IBM Heron (heavy-hex) and IonQ Forte (all-to-all). 
The dominant compiler bottleneck is strongly circuit-class dependent: Routing accounts for up to 60\% of relative fidelity loss in search-class circuits, while synthesis dominates Hamiltonian simulation workloads. 
Early synthesis choices amplify or compress downstream routing overhead depending on circuit connectivity. 
SDK rankings at diagnostic optimization level (\texttt{opt=0}) reverse at production levels (\texttt{opt=2}) for deep circuits, showing that stagewise diagnostics and production results answer different questions. 
HBR correctly predicts SDK rank ordering across noisy simulations (8 circuits~$\times$~3 SDKs~$\times$~2 tiers) and real IBM Fez hardware executions, revealing stage-specific bottlenecks that are not observable through aggregate compiler benchmarks.
\end{abstract}

\section{Introduction}
\label{sec:intro}

In the NISQ era~\cite{preskill:quantum:2018}, quantum compilers sit between an algorithm's abstract circuit description and what ultimately executes on noisy hardware. The compilation pipeline can materially change a circuit's gate count, depth and qubit interactions, and therefore its expected fidelity on a given device~\cite{zhu:arxiv:2025}. While different SDKs implement this pipeline differently, their output is commonly judged using aggregate post-transpilation metrics such as total two-qubit gates, circuit depth or compile time. However, aggregate metrics alone do not explain \emph{why} a compiler performs better or worse, nor do they indicate \emph{which part} of a pipeline should be improved.

Most compilation flows, even when not explicitly exposed by an SDK, can be abstracted into three conceptual stages: \emph{synthesis}, \emph{basis translation}, and \emph{routing}~\cite{zhu:arxiv:2025}. \emph{Synthesis} decomposes high-level gates (e.g., Toffoli/MCX) into elementary single-qubit (1Q) and two-qubit (2Q) primitive gates and applies local simplifications. \emph{Basis translation} rewrites that intermediate gate set into the backend-native instruction set (ISA). \emph{Routing} inserts SWAPs~\cite{li:asplos:2019} and chooses qubit placements to satisfy the device coupling constraints, which vary strongly across hardware topologies. Each stage changes gate count and depth and thereby exposes itself to gate errors and decoherence. As a result, the dominant bottleneck is often \emph{circuit-class dependent}: a circuit that is synthesis-heavy may be insensitive to routing, while a circuit with dense long-range interactions may be dominated by SWAP overhead on sparse coupling graphs.

This motivates a diagnostic question that existing benchmarks largely cannot answer: \textit{for a given circuit class and target backend, \underline{where} is fidelity being lost during compilation?} Benchmarks such as Benchpress~\cite{nation:ncs:2025} are valuable for comparing end-to-end outcomes, but they treat the compiler largely as a black box and report only aggregate overhead. Without phase-specific attribution, observations such as ``SDK~X wins by $k$ gates'' are difficult to act upon. A compiler developer cannot tell whether to improve synthesis rules, basis translation or routing heuristics, and an SDK user cannot tell whether changing optimization settings is likely to help on their circuit family. Further, modern production compilation pipelines often interleave transformations across stages~\cite{javadi:arxiv:2024} (especially at higher optimization levels), making it challenging to reason about how early-stage decisions influence later routing and optimization.

In this paper, we address this gap with \emph{HBR decomposition}, our proposed per-phase fidelity attribution methodology that quantifies the aggregate fidelity cost attributable to three stages: \underline{H}igh-level structural decomposition/synthesis (H), \underline{B}asis translation (B), and \underline{R}outing (R). Our goal is not to predict the absolute hardware fidelity for deep circuits, but rather to provide an actionable, phase-resolved diagnostic that preserves relative comparisons across SDKs and reveals which compilation stages dominate fidelity loss for a given circuit class. We evaluate three widely used SDKs: \textit{Qiskit}~\cite{javadi:arxiv:2024}, \textit{PennyLane}~\cite{bergholm:arxiv:2022}, and \textit{TKET}~\cite{sivarajah:qst:2020}, on eight representative algorithms across two contrasting quantum computing backends: an \textit{IBM Heron-class}~\cite{ibm:arxiv:2024} superconducting backend with heavy-hex connectivity,
and an \textit{IonQ Forte-class}~\cite{chen:quantum:2024} trapped-ion backend with all-to-all connectivity. This contrast allows us to separate synthesis/translation effects from topology-driven routing overhead, and to highlight when compiler behavior changes across diagnostic versus production optimization tiers.

We make the following specific contributions in this paper:
\begin{enumerate}[leftmargin=*,nosep]
\item We introduce HBR decomposition to attribute \emph{relative fidelity loss} to three compilation stages: High-level decomposition/synthesis (\textbf{H}), Basis translation (\textbf{B}) and Routing (\textbf{R}), and describe how we instrument stage boundaries across SDKs (\S\ref{sec:framework}).
\item We define a \textit{CX-equivalent (CXeq) accounting unit} to normalize 2Q costs across SDKs/backends (\S\ref{sec:framework:cx-equiv}), and use a lightweight analytic model to convert per-stage gate/depth overheads into \textit{additive log-fidelity contributions} (\S\ref{sec:model}).
\item Using Qiskit, PennyLane, and TKET on IBM heavy-hex and IonQ all-to-all topologies, we show that \textit{dominant bottlenecks} are circuit-class dependent and that \textit{early synthesis choices} can amplify or compress routing overhead. We further demonstrate that diagnostic \texttt{opt=0} winners can reverse at production \texttt{opt=2}.
We further \textit{validate} that the model preserves SDK rank ordering across IBM FakeFez noisy simulation, IonQ Forte-1 noisy simulation and IBM Fez real hardware of BV ($n=12$, deterministic) and Grover ($n=4$, deep), and assess \textit{sensitivity} to calibration perturbations and routing stochasticity (\S\ref{sec:experiments}).
\end{enumerate}
Besides these, \S\ref{sec:related} contextualises related work,
\S\ref{sec:discussion} discusses future directions and limitations, and \S\ref{sec:conclusion} concludes the work.

\section{Related Work}
\label{sec:related}

\subsection{Compiler benchmarks}
\textit{IBM's Benchpress}~\cite{nation:ncs:2025} benchmarks seven quantum SDKs on aggregate metrics across different circuit types and topologies.
\textit{SupermarQ}~\cite{tomesh:hpca:2022} defines feature vectors (entanglement, parallelism, measurement) to characterize circuit difficulty for hardware prediction.
\textit{QASMBench}~\cite{li:tqci:2023} provides a low-level QASM benchmark suite for NISQ evaluation with aggregate gate density and lifespan metrics.
\textit{QED-C}~\cite{lubinski:tqe:2023} defines
application-oriented benchmarks measuring algorithm-level correctness on real hardware. However, these benchmarks focus on functional correctness rather than identifying the architectural source of
compiler overhead.

Furthermore, meta-compilation frameworks like \textit{MQT Predictor}~\cite{quetschlich:tqc:2025} automate SDK selection but operate as black boxes. HBR is complementary by providing the white-box diagnostics needed to explain \emph{why} an SDK performs differently at the synthesis or routing levels.
Beyond software benchmarks, \textit{Quantum characterization, verification, and validation (QCVV)} provides a broad toolbox for benchmarking quantum \emph{devices} and their noise processes, ranging from qubit-level characterization to scalable benchmarking protocols~\cite{hashim:prxq:2025}. These efforts are
complementary: they assess hardware behavior or end-to-end execution quality, whereas HBR targets \emph{compiler-internal} attribution, i.e., identifying which compilation phase (H/B/R) is the dominant
source of fidelity loss for a given circuit class and topology.

\subsection{Pipeline-level profiling and stage interleaving}
\textit{Zilk et~al.\@}~\cite{zilk:ISVLSI:2025} profile Qiskit's \textit{cProfile} execution to identify which compiler passes dominate \emph{classical} compilation time. Our work is orthogonal: we profile
\emph{quantum fidelity loss} per stage rather than CPU time, and we do so across three SDKs and two hardware connectivity extremes. Importantly, production pipelines at higher optimization levels interleave and reorder transformations (e.g., optimization and translation/resynthesis after routing)~\cite{javadi:arxiv:2024}, which makes phase boundaries non-trivial and motivates our separation between diagnostic (\texttt{opt=0}) and production (\texttt{opt=2}) analyses.

\subsection{Noise models}
While metrics like \textit{Estimated Success Probability (ESP)} or \textit{Map-Fidelity}~\cite{mills:arxiv:2020} provide single-value hardware readiness estimates, they lack the stage-wise attribution required for compiler diagnostics. The independent depolarizing channel composition underlying our per-phase formula is standard~\cite{nielsen:qcqi:2000} and was used in the \textit{Cross-entropy Benchmarking (XEB)} analysis~\cite{arute:nature:2019}.
Georgopoulos et~al.\@~\cite{georgopoulos:pra:2021} model edgewise heterogeneous noise and provide the per-qubit idle relaxation formula we adopt for $F_{T_2}$, which we validate using hardware rank
ordering.
Recently, Escofet et~al.\@~\cite{escofet:QST:2025} proposed a qubitwise iterative fidelity algorithm that tracks per-qubit depolarization state through each gate layer, but it requires a circuit-wise tuning parameter $p_{\text{ent}} \in [0,1]$ controlling assumed entanglement.
In contrast, HBR uses lightweight analytic modeling specifically to obtain \emph{additive, phasewise} log-fidelity deltas from gate/depth changes at stage boundaries, emphasizing actionable diagnostics and rank preservation over absolute fidelity prediction.

\subsection{Routing}
\textit{SABRE}~\cite{li:asplos:2019} is the dominant qubit routing heuristic in Qiskit. TKET uses \texttt{Graph\-Placement+Routing\-Pass}~\cite{cowtan:tqc:2019,sivarajah:qst:2020}, which is deterministic
unlike SABRE's stochastic approach.
Topology-aware benchmarks~\cite{mills:arxiv:2020} highlight the variance in routing cost across different layouts, motivating our evaluation on both sparse heavy-hex and all-to-all connectivity.
A noise-adaptive mapping~\cite{murali:asplos:2019} selects qubit placement using calibration data, whereas our attribution framework uses bulk median parameters so differences in HBR breakdown reflect compiler behavior rather than hardware selection effects.

Related routing research also argues that routing decisions should account for \emph{downstream} optimization. E.g., \textit{NASSC} observes that inserted SWAPs differ in their basis-gate cost after subsequent optimizations and proposes optimization-aware routing accordingly~\cite{liu:hpca:2022}.
Our work is complementary: rather than proposing a new router, HBR quantifies \emph{how much} of the total fidelity loss is attributable to routing versus synthesis/translation, and exposes when early synthesis choices amplify (or compress) routing overhead.

\section{The HBR Framework}
\label{sec:framework}

\begin{figure}[t]
\centering
\includegraphics[width=1\columnwidth]{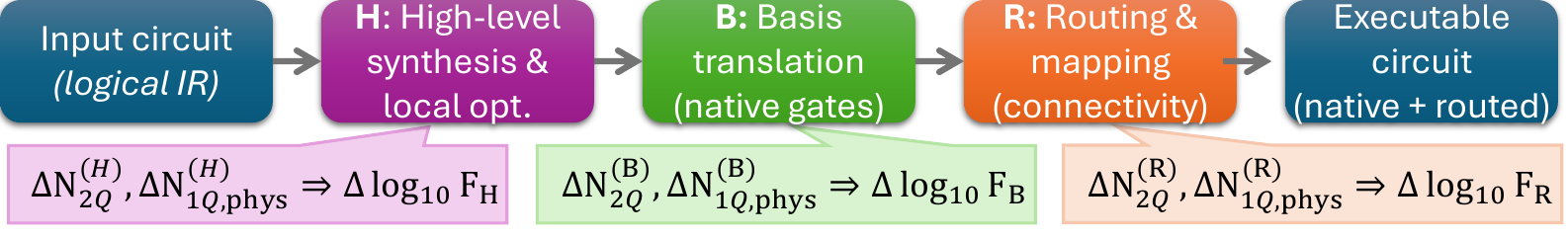}
\caption{
HBR compilation pipeline. Each stage $X\in\{H,B,R\}$ is metered by changing physical gate counts, which are mapped to $\Delta\log_{10}(F_X)$.}
\label{fig:hbr-pipe}
\end{figure}

We model compilation as a three-stage pipeline: \emph{H}igh-level synthesis, \emph{B}asis translation and \emph{R}outing, and instrument each stage to measure its independent contribution to circuit cost. Although most quantum SDKs do not expose these as explicit boundaries, every compiler performs these transformations in some form.

\subsection{The HBR Pipeline}
\label{sec:framework:pipeline}
The HBR pipeline decomposes compilation into three conceptual stages:
\begin{enumerate}[leftmargin=*,nosep]
    \item \textit{\textbf{H}: High-level synthesis and local optimization.} Converts algorithmic and composite operations into a lower-level gate model while applying algebraic simplifications and minor optimizations. The output may still contain non-native 2Q gates.
    \item \textit{\textbf{B}: Basis translation.} Rewrites the circuit into the backend's native gate set without changing the logical qubit topology. This stage should not introduce connectivity-driven SWAPs.
    \item \textit{\textbf{R}: Routing and physical mapping.} Maps logical qubits onto the backend coupling graph and inserts SWAPs (or equivalent move operations) to satisfy connectivity constraints, optionally followed by post-routing cleanup.
\end{enumerate}
Fig.~\ref{fig:hbr-pipe} summarizes the HBR pipeline and the per-stage quantities we instrument, which form the basis of the stagewise overhead analysis in \S\ref{sec:results:hbr}.

\subsection{HBR Instrumentation by SDK}
\label{sec:framework:hbr}

\begin{figure}[t]
\centering
\includegraphics[width=1\columnwidth]{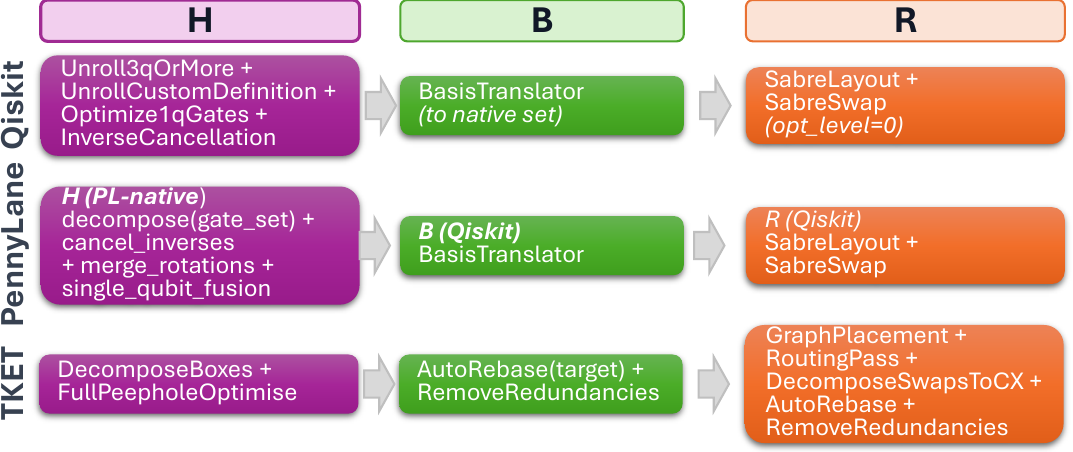}
\caption{
HBR compilation pipeline for each SDK. Operations assigned to each conceptual H/B/R stage per SDK are shown.}
\label{fig:sdk-pipe}
\end{figure}

No quantum SDK exposes a clean HBR boundary. We therefore implement SDK-specific pass sequences that operationally approximate the conceptual H/B/R stages and measure the changing circuit configuration added at each stage.
Fig.~\ref{fig:sdk-pipe} summarizes the instrumentation map used for Qiskit, PennyLane and TKET.

\subsubsection{Qiskit}
We use Qiskit's staged compilation to separate H/B/R stages. The operations present in each stage are given in Fig.~\ref{fig:sdk-pipe}.
Routing is executed with \texttt{opt\_level=0} in \textit{Sabre} to isolate routing overhead from additional heuristic tuning.

\subsubsection{PennyLane}
PennyLane uses a native H-stage (decomposition and local algebraic simplifications), while its B and R stages reuse Qiskit passes and are metered separately.
Consequently, PennyLane-only effects appear in H.

\subsubsection{TKET}
TKET does not provide an exposed HBR boundary. We manually factor its compilation into H/B/R stages as shown in Fig.~\ref{fig:sdk-pipe}, including a post-routing cleanup to remove redundancies.

To prevent \texttt{CircBox} con\-version failures on nested composite operations (e.g., \texttt{PhaseEstimation}, \texttt{Pauli\-Evolution\-Gate}), all input circuits are pre-flattened using Qiskit format-conversion passes \texttt{Unroll3q\-OrMore} and \texttt{Unroll\-Custom\-Definitions}. These passes are used strictly for structural normalization, without optimization.

\subsection{Stage Boundary Definitions}
\label{sec:framework:boundary}

The H/B boundary is not uniform across SDKs because it depends on the Intermediate Representation (IR) and compilation strategy. E.g., Qiskit's H stage expands composites primarily into 1Q/2Q primitives while leaving some non-native 2Q forms to be canonicalized in B, whereas PennyLane may decompose certain high-level operators (e.g., multi-controlled gates or \texttt{ApproxTimeEvolution}) earlier and then apply local optimizations within H.

Because the boundary between H and B is SDK-dependent, we adopt the following consistency rules: \textit{(i)~Within-SDK Interpretation:} Per-stage H and B costs are reported as-in and are comparable only \emph{within} the same SDK; \textit{(ii)~Cross-SDK Comparability:} The combined pre-routing cost, $\Delta\log_{10}(F_{H+B})$, is the appropriate cross-SDK comparison metric for total compilation overhead before routing; and \textit{(iii)~Boundary-sensitive Algorithms:} In \S\ref{sec:results:hbr}, H\% and B\% values are marked with $\star$ for algorithm families whose decomposition can shift across the H/B boundary (e.g., QDRIFT and Trotterized simulation).

\subsection{CX-Equivalent Gate Costing}
\label{sec:framework:cx-equiv}

Comparison across SDKs and backends requires a common unit of 2Q cost. We adopt \emph{CX-equivalent (CXeq)} as a universal accounting unit for two-qubit operations.

\begin{table}[t]
\centering
\caption{CX-equivalent (CXeq) costing rules.}
\label{tbl:cx-equiv}
\begin{tabular}{l|c|l}
\hline
\bf Gate & \bf CXeq & \bf Ref \\
\hline\hline
CX, CZ, MS/XX & 1 & \cite{shende:tcad:2006,maslov:npj:2018} \\
CP($\theta$), CRZ($\theta$) & 2$^\ast$ & \cite{shende:tcad:2006} \\
RXX, RYY, RZZ & 2$^\ast$ & \cite{vatan:pra:2004} \\
SWAP & 3 & \cite{shende:tcad:2006} \\
\hline
\multicolumn{3}{l}{\footnotesize $^\ast$2~CX lower bound for non-maximally-entangling} \\
\multicolumn{3}{l}{\footnotesize \phantom{$^\ast$}gates~\cite{shende:tcad:2006}. CP($\pi$) = CZ = 1~CXeq.} \\
\end{tabular}
\end{table}

CXeq serves two purposes:
\textit{(i)~SDK Comparability:} Different SDKs target different native 2Q gates (e.g., CX/CZ for superconducting devices, MS/XX for trapped-ion devices). CXeq makes HBR gate counts directly comparable across SDKs.
\textit{(ii)~Backend Comparability:} IBM Heron's native 2Q gate (CZ) and IonQ Forte's native 2Q gate (MS) both cost 1~CXeq. Routing overhead is accounted using SWAPs, costed at 3~CXeq each (consistent with a 3-CX decomposition before execution).

CX equivalence is used only for \textit{phasewise gate counting}. Simulation and hardware validation (\S~~\ref{sec:model:variants}) use per-edge and per-gate calibration parameters so that those results evaluate the noise model directly, not this accounting convention.

\subsection{Virtual Gates Exclusion}
\label{sec:framework:virtual}

On IBM superconducting backends, \texttt{rz} is implemented as a virtual frame update with zero duration and (to first order) zero physical error~\cite{mckay:pra:2017}. A similar property holds for IonQ Forte, where \texttt{rz} is implemented as a virtual phase advance in the Raman laser frame~\cite{chen:quantum:2024,maslov:npj:2018}. Including such virtual operations can inflate reported $N_{1Q}$ by $2$--$4\times$ without reflecting additional physical error. All reported $N_{1Q}$ counts therefore exclude virtual gates (\texttt{rz}, \texttt{p}, \texttt{u1}, \texttt{GlobalPhase}).

\section{Fidelity Model}
\label{sec:model}

We use an analytical fidelity model to attribute compilation overhead to the H/B/R stages, and estimate end-to-end circuit fidelity under a simple, transparent noise abstraction. The model is intentionally lightweight: it is designed to preserve \emph{relative} SDK comparisons and stagewise attribution, not to exactly reproduce hardware outcomes under correlated and coherent noise (assumptions in \S\ref{sec:model:assumptions}).

\subsection{Analytical Formula}
\label{sec:model:formula}

\subsubsection{Per-Phase Attribution}
We attribute compilation-induced fidelity loss to each HBR stage by converting stage-added gate counts into an additive contribution in $\log_{10}$ fidelity space.
\begin{align}
\label{eq:phase-fidelity}
    \Delta\!\log_{10}\!(F_X)
    &= \Delta N_{2Q}^{(X)} \cdot \log_{10}(1-p_{2Q})\nonumber\\
    &\quad
    + \Delta N_{1Q,\text{phy}}^{(X)} \cdot \log_{10}(1-p_{1Q})
\end{align}
where $X \in \{H,B,R\}$ denotes a compiler stage, and $\Delta N_{2Q}^{(X)}$ and $\Delta N_{1Q,\text{phy}}^{(X)}$ are the gate counts changed at stage $X$ relative to the preceding stage.
Eqn.~\ref{eq:phase-fidelity} is exact under independent depolarizing noise~\cite{nielsen:qcqi:2000}, i.e., each gate contributes a multiplicative factor $(1-p)$ in linear fidelity, which becomes additive in $\log_{10}$ space, enabling clean stagewise decoupling.

At the H stage, 2Q primitives (e.g., \textsc{cx}, \textsc{tk1}-based decompositions) may not yet be expressed in the backend-native basis. We apply the backend-median $p_{2Q}$ uniformly to all H-stage 2Q gates as a \emph{counting proxy}. The absolute $\Delta\!\log_{10}(F_H)$ is therefore not a hardware-accurate fidelity prediction, but \emph{differences} between SDKs at H remain meaningful as relative overhead. The combined $\Delta\!\log_{10}(F_{H+B})$ is the cross-SDK metric per \S\ref{sec:framework:boundary}. The B stage then captures the remaining overhead of translating synthesized gates into the native hardware basis.

\begin{figure}[t]
\centering
\includegraphics[width=1\columnwidth]{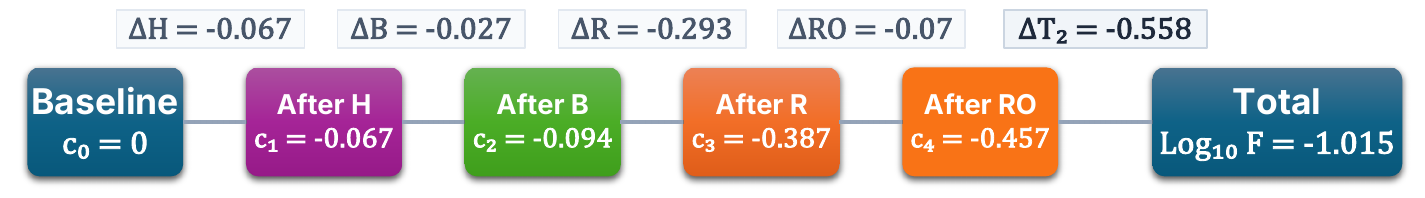}
\caption{Example of cumulative contributions to $\log_{10}(F)$ from H/B/R stages and compiler-agnostic terms (readout and $T_2$ idle decoherence) for QFT at n=10.}
\label{fig:waterfall}
\end{figure}

\subsubsection{Total Circuit Fidelity}
We obtain an end-to-end fidelity estimate by aggregating the H, B, and R stage contributions and then adding compiler-agnostic readout and idle-decoherence terms.
\begin{align}
\label{eq:total-fidelity}
    \log_{10}(F)
    &= \sum_{X \in \{H,B,R\}} \Delta\!\log_{10}(F_X) \nonumber\\
    &\quad + n \cdot \log_{10}(1 - p_{\text{RO}})
    + \log_{10}(F_{T_2})
\end{align}
where $n$ is the number of measured qubits and $p_{\text{RO}}$ is the per-qubit readout error.
The readout term $n \cdot \log_{10}(1-p_{\text{RO}})$ is compiler-agnostic and is therefore excluded from phasewise percentage breakdowns, but included here to improve the absolute fidelity estimate.

We model additional decoherence from idle periods using an effective $T_2$ penalty on qubits not actively undergoing gate operations:
\begin{align}
T_{\text{idle}} &= n_{\text{qubits}} \cdot t_{\text{circuit}} - T_{\text{active}} \\
T_{\text{active}} &= N_{2Q}\,\tau_{2Q}\cdot 2 + N_{1Q,\text{phys}}\,\tau_{1Q} \\
t_{\text{circuit}} &= D_{2Q}\,\tau_{2Q} + D_{1Q}\,\tau_{1Q}
\end{align}
where $D_{2Q}$ is the two-qubit layer depth (critical path) and $D_{1Q}=D_{\text{tot}}-D_{2Q}$ is the remaining one-qubit layer depth~\cite{tannu:asplos:2019}. Thus, two circuits with identical $N_{2Q}$ but different depth can incur different $T_{\text{idle}}$ and hence different decoherence.
Each 2Q gate occupies two qubits for duration $\tau_{2Q}$, and each physical 1Q gate occupies one qubit for duration $\tau_{1Q}$.
Under exponential dephasing, the resulting idle contribution is~\cite{georgopoulos:pra:2021}:
\begin{equation}
\label{eq:decoherence}
    \log_{10}(F_{T_2}) = -\left(\frac{T_{\text{idle}}}{T_2}\right)\log_{10}e
\end{equation}
In linear space,
$F = \Big(\prod_{X \in \{H,B,R\}} F_X\Big)\cdot F_{\text{RO}} \cdot F_{T_2}$,
and Eqn.~\ref{eq:total-fidelity} is its $\log_{10}$-space equivalent.

Fig.~\ref{fig:waterfall} illustrates this additive decomposition on QFT circuit at n=10, showing cumulative $\Delta\log_{10}(F)$ contributions from H, B and R compilation stages and the compiler-agnostic readout and idle-decoherence terms.

\subsection{Analytical Model Variants}
\label{sec:model:variants}
The product model (Eqs.~\ref{eq:phase-fidelity},~\ref{eq:total-fidelity}) is used in two configurations (Table~\ref{tbl:model-variants}). The \emph{attribution} configuration uses backend-median error rates to isolate compiler behavior. The \emph{validation} configuration uses per-edge/per-qubit parameters extracted from the target noise model.

\begin{table}[t]
\caption{Model configurations. \emph{Uniform} uses backend-median scalars. \emph{Per-edge} uses parameters extracted from \texttt{NoiseModel.from\_backend()}.}
\label{tbl:model-variants}
\centering
\begin{tabular}{llll}
\hline
\bf Configuration & \bf Error rates & \bf $T_2$ idle & \bf Used for \\
\hline
\hline
\textbf{Attribution} & Uniform & Yes & HBR gaps (Table~\ref{tbl:hbr-deep}) \\\hline
\textbf{Validation} & Per-edge & Context$^\dagger$ & Simulators (Table~\ref{tbl:sim-validation}), IBM Fez (Table~\ref{tbl:hw-validation}) \\
\hline
\multicolumn{4}{l}{\footnotesize $^\dagger$FakeFez: $T_2$ excluded (Aer per-gate channels include $T_1$/$T_2$ effects).} \\
\multicolumn{4}{l}{\footnotesize \phantom{$^\dagger$}IBM Fez hardware: per-qubit $T_2$ included from job calibration snapshot.}
\end{tabular}
\end{table}

\subsubsection{Attribution configuration}
We include $T_2$ idle decoherence because thermal relaxation and dephasing act continuously on real hardware, including on idle qubits. This contribution is not captured by gate error rates derived from randomized benchmarking alone~\cite{hashim:prxq:2021}.

\subsubsection{Validation configuration}
For simulation-based validation, we use per-edge and per-qubit error rates and readout parameters extracted from \texttt{Noise\-Model.from\_backend(FakeFez)}. We exclude the explicit $T_2$ idle penalty because Aer attaches thermal relaxation channels to gate operations (capturing in-gate $T_1$/$T_2$ decay), but does not model relaxation on idle qubits unless explicit \texttt{Delay} instructions are introduced through scheduling~\cite{qiskit:aer:2023}. Since validation circuits are transpiled to match topology without scheduling passes, Aer treats idle intervals as perfectly coherent. Adding an idle penalty would therefore be inconsistent with the simulator's effective noise semantics.

\subsection{Model Assumptions}
\label{sec:model:assumptions}

\begin{table}[t]
\centering
\caption{Model assumptions and their impact on SDK rankings.}
\label{tbl:assumptions}
\begin{tabular}{ll L{4cm}}
\hline
\bf Assumption & \bf Reality & \bf Impact \\
\hline\hline
Independent Markovian noise & ZZ $\approx$20-60\,kHz crosstalk~\cite{kandala:prl:2021} & Rankings unaffected \\\hline
Scalar $T_2$/backend & Qubit-specific & If only SDKs route very differently \\\hline
No leakage to $|2\rangle$ & Per-gate leakage $\lesssim$1\% \cite{wood:pra:2018,varbanov:npjqi:2020} & SDK-independent \\\hline
Depolarizing only & Coherent $O(\varepsilon^2 N^2)$ \cite{wallman:pra:2016,sanders:pra:2015} & Only absolute $F$ affected \\\hline
Uniform $p_{2Q}$ & Edges vary $1.2$--$5\times 10^{-3}$ & Heterogeneous model used for validation \\
\hline
\end{tabular}
\end{table}

Because our goal is to \emph{rank} compiler overhead rather than predict absolute hardware fidelity, the absolute $\log_{10}(F)$ values are not expected to match hardware exactly, but the SDK \emph{rankings} are robust under these simplifications.
Under correlated noise, $F(E) \leq F(E_H) \cdot F(E_B) \cdot F(E_R)$ with equality only when correlations vanish to $0$.
Coherent errors can accumulate as $O(\varepsilon^2 N^2)$~\cite{wallman:pra:2016,sanders:pra:2015}, whereas stochastic depolarizing errors scale as $O(\varepsilon N)$. This primarily impacts absolute fidelity for deep circuits, but typically does not change rank ordering across SDKs under the same benchmark family.
Table~\ref{tbl:assumptions} summarizes the modeling assumptions and their expected impact on SDK rank ordering.

\section{Experimental Results}
\label{sec:experiments}

We evaluate our HBR framework for three SDKs in two tiers to separate \emph{Tier~1 (diagnostic)} per-stage attribution from \emph{Tier~2 (production)} end-to-end performance.
Tier~1 uses HBR at \texttt{opt=0} to enforce strict H/B/R boundaries and attribute overhead to synthesis, basis translation and routing, while Tier~2 uses production pipelines (\texttt{opt=2}/default) where passes may interleave and only total outcomes are compared.
Across both tiers we benchmark eight circuits on IBM heavy-hex and IonQ all-to-all, study scaling over $n\in[3,20]$, validate rank ordering using noisy simulators and IBM Fez hardware, and perform robustness checks for calibration and routing stochasticity.

\subsection{Setup}
\label{sec:setup}

\subsubsection{Benchmarks and fixed parameters}
\label{sec:setup:algos}
We evaluate eight representative circuits (Table~\ref{tbl:algorithms}) chosen so that each stresses a different compilation bottleneck: high-level synthesis (deep decompositions and local rewrites), basis translation or topology-induced routing overhead.

\textit{Grover} uses the standard iteration count $k=\mathrm{round}(\pi/4\cdot \sqrt{2^n})$ (thus $k{=}25$ at $n{=}10$) and marks the state $\min(42,2^n{-}1)$.
Hamiltonian simulation workloads (\textit{QDRIFT}, \textit{Trotter} and \textit{QPE} unitary) use the Transverse-Field Ising Model (TFIM) with $(J,h,t)=(1.0,0.5,1.0)$. QDRIFT uses $N=(\lambda t)^2/(2\varepsilon)=1960$ with $\lambda=14$ and $\varepsilon=0.05$, averaged over five channel seeds ($42$--$46$).
For \textit{Trotter-Suzuki}, the step count is fixed \emph{before} compilation and identical across SDKs. It is chosen from the set in Table~\ref{tbl:algorithms} to balance algorithmic approximation error against circuit noise.
\textit{BV} uses a fixed secret string generated with seed=42 and a star-topology oracle that maximizes routing stress on heavy-hex.
\textit{QAOA} is a $p=1$ MaxCut instance on a 3-regular graph (seed=42) with 15 edges and fixed angles. Additional details are in Table~\ref{tbl:algorithms}.

\begin{table}[t]
\centering
\caption{Benchmark circuits used in the evaluation  at $n{=}10$.}
\label{tbl:algorithms}
\begin{tabular}{l | l r}
\hline
\textbf{Circuit} & \textbf{Bottleneck} & \textbf{Key parameters (fixed)} \\
\hline\hline
\bf GHZ~\cite{ghz:bells:1989}
 & Synthesis (H) & H+CX entangled chain \\
\bf Grover~\cite{grover:stoc:1996}
 & Synthesis+Routing & MCX oracle, 25 iterations \\
\bf QDRIFT~\cite{campbell:prl:2019}
 & Synthesis (H/B) & TFIM, $N{=}1960$, 5 seeds \\
\bf QFT~\cite{coppersmith:ibm:1994}
 & Routing & $O(n^2)$ controlled rotations \\
\bf QPE~\cite{kitaev:arxiv:1995}
 & Synthesis+Phase est. & TFIM, 4 ancilla, reps$=4$ \\
\bf Trotter~\cite{suzuki:jmp:1991}
 & H/B boundary & TFIM, steps $\in\{1,2,4,6,8,10\}$ \\
\bf BV~\cite{bernstein:sjc:1993}
 & Routing & Star-oracle (4 CNOTs) \\
\bf QAOA~\cite{farhi:arxiv:2014}
 & Synthesis (H) & MaxCut $p{=}1$, 3-regular, 15 edges \\
\hline
\end{tabular}
\end{table}

\subsubsection{SDKs and Backends}
\label{sec:setup:sdks}

\begin{table}[t]
\centering
\caption{Hardware noise parameters.}
\label{tbl:noise-params}
\begin{tabular}{l|rr}
\hline
\bf Parameter & \bf IBM Heron r2~\cite{ibm:arxiv:2024} & \bf IonQ Forte~\cite{chen:quantum:2024} \\
\hline\hline
$p_{2Q}$ & $3\times10^{-3}$ & $4\times10^{-3}$ \\
$p_{1Q}$ & $3\times10^{-4}$\ & $2\times10^{-4}$ \\
$p_{\text{RO}}$ & $1.6\times10^{-2}$~ & $7\times10^{-3}$ \\
$\tau_{2Q}$ & 68\,ns (CZ) & 600\,$\mu$s (MS) \\
$\tau_{1Q}$ & 32\,ns (SX) & 10\,$\mu$s \\
$T_2$ & 79\,$\mu$s & $\approx$200\,ms\tablefootnote{Operational estimate. Fieldfree $T_2 \approx 1$\,s~\cite{bruzewicz:apr:2019}, reduced by AC Stark shifts during gate execution.} \\
Topology & Heavy-hex 156q (sparse) & all-to-all 36q \\
\hline
\end{tabular}
\end{table}

All experiments use \textit{Qiskit (v2.3.0)}, \textit{PennyLane (v0.44.0)} and \textit{TKET (v2.15.0)}, with Python (v3.14.0).
We evaluate two contrasting hardware topologies: an \textit{IBM Heron-class} heavy-hex backend (sparse connectivity) and an \textit{IonQ Forte-class} trapped-ion backend (all-to-all connectivity).
Table~\ref{tbl:noise-params} summarizes the scalar noise parameters used by the attribution model and the native gate durations used to compute the $T_2$ idle term.

\subsubsection{Fairness Controls}
\label{sec:setup:fairness}
We apply the following controls to ensure that observed differences reflect compiler behavior rather than experimental variance:
\begin{itemize}[leftmargin=*,nosep]
\item \textbf{Determinism and budget.} Default results use seed=42 and shots=8{,}192 at $n{=}10$. Scaling is evaluated over $n\in[3,20]$ with a 300\,s timeout.
\item \textbf{Diagnostic routing tier \texttt{(opt=0)}.} Routing is performed at \texttt{opt\_level=0} to isolate raw SWAP insertion cost. Higher optimization levels apply post-routing cleanup (e.g., commutation analysis and SWAP cancellation), so \textit{opt$\ge2$} represents a different question (production performance) and is evaluated separately in \S\ref{sec:crosstier}.
\item \textbf{Routing stochasticity.} H and B are deterministic. Only Qiskit's SABRE routing is stochastic. We quantify variance using 25 transpiler seeds (16, 18, 20, \ldots, 64). TKET routing is deterministic ($\sigma=0$ across seeds).
\item \textbf{Backend topology.} The IBM coupling map is based on \texttt{ibm\_fez} (156 qubits, heavy-hex, hardcoded from FakeFez), while IonQ Forte assumes all-to-all connectivity (36 qubits) (Table \ref{tbl:noise-params}).
\end{itemize}
All code and experiment scripts are available at \url{https://github.com/dream-lab/quantum-hbr}.

\subsection{Diagnostic Results (\texttt{opt=0})}
\label{sec:results}

\subsubsection{Overview}
\label{sec:results:overview}

\begin{figure}[t]
  \centering
  \includegraphics[width=0.75\columnwidth]{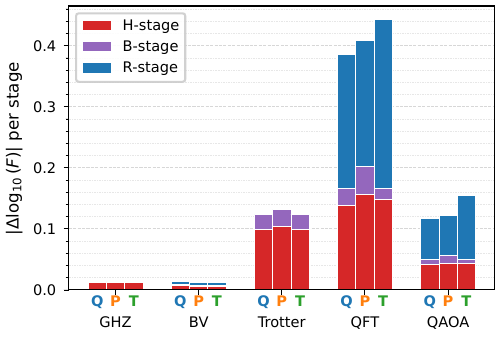}
  \caption{Per-phase fidelity attribution for shallow algorithms across three SDKs on IBM. All gaps $<0.5$ decades.}
  \label{fig:hbr-shallow}
\end{figure}

At \texttt{opt=0}, the HBR instrumentation enforces strict stage boundaries (Fig.~\ref{fig:hbr-pipe}--\ref{fig:sdk-pipe}), allowing us to attribute overhead to H/B/R without cross-stage interleaving.
Across the eight benchmarks at $n{=}10$, five circuits show inter-SDK gaps below our comparability threshold ($<0.5$ decades in $\Delta\log_{10}(F_{H+B+R})$) on IBM (Fig.~\ref{fig:hbr-shallow}). Total $\log_{10}F$ gap exceeds the H+B+R sum because $T_2$ idle decoherence amplifies depth differences.
Three circuits exhibit substantial compiler gaps and are analyzed in detail:\textbf{Grover} (TKET wins by 7.2d H+B+R on IBM, driven by both H+B synthesis and routing), \textbf{QPE} (TKET wins by 6.0d, primarily routing, with 7,995 SWAPs vs.\ Qiskit's 11,631), and \textbf{QDRIFT} (TKET wins by 1.5d H+B over Qiskit via H-stage synthesis, 1,546 2Q gates vs.\ 2,507). These are diagnostic findings. Production-tier comparisons appear in \S\ref{sec:crosstier}.

\subsubsection{Full HBR Attribution}
\label{sec:results:hbr}

\begin{table}[!t]
\centering
\def\thickhline{\noalign{\hrule height1pt}}

\caption{Per-phase fidelity attribution ($n{=}10$, seed=42, \texttt{opt=0}) on IBM Heron (heavy-hex) and IonQ Forte (all-to-all). Values are $\Delta\log_{10}F$ decades (negative = fidelity cost). Phase percentages over $|\Delta F_{H\!+\!B\!+\!R}|$.}
\label{tbl:hbr-deep}
\begin{tabular}{l|l|l|r|r|r|r|r|r|r}
\hline
\bf Algo. & \bf SDK & \bf Back. & $\Delta F_H$ & $\Delta F_B$ & $\Delta F_{H\!+\!B}$ & $\Delta F_R$ & \bf H\% & \bf B\% & \bf R\% \\
\hline\hline
\multirow{6}{*}{\textbf{Grover}}
 & TKET      & IBM  & $-$31.9 & $-$3.9  & $-$35.7 & $-$50.2 & 37.1 & 4.5 & 58.4 \\
 & TKET      & IonQ & $-$40.6 & $+$1.4  & $-$39.2 &     0.0 & 96.7 & 3.3 &  0.0 \\ \cmidrule{2-10}
 & Qiskit    & IBM  & $-$33.4 & $-$4.2  & $-$37.6 & $-$55.6 & 35.8 & 4.5 & 59.7 \\
 & Qiskit    & IonQ & $-$41.5 & $+$0.2  & $-$41.3 &     0.0 & 99.4 & 0.6 &  0.0 \\\cmidrule{2-10}
 & PennyLn. & IBM  & $-$57.0 & $-$16.4 & $-$73.4 & $-$88.7 & 35.1 & 10.1 & 54.7 \\
 & PennyLn. & IonQ & $-$71.8 &   0.0   & $-$71.8 &     0.0 & 100.0 & 0.0 & 0.0 \\
\thickhline
\multirow{6}{*}{\textbf{QPE}}
 & TKET      & IBM  & $-$13.8 & $-$2.0 & $-$15.9 & $-$12.5 & 48.7 & 7.2 & 44.1 \\
 & TKET      & IonQ & $-$17.6 & $+$0.5 & $-$17.2 &     0.0 & 97.4 & 2.6 &  0.0 \\\cmidrule{2-10}
 & Qiskit    & IBM  & $-$13.5 & $-$2.7 & $-$16.2 & $-$18.2 & 39.1 & 7.9 & 52.9 \\
 & Qiskit    & IonQ & $-$17.0 & $-$0.8 & $-$17.8 &     0.0 & 95.6 & 4.4 &  0.0 \\\cmidrule{2-10}
 & PennyLn. & IBM  & $-$15.5 & $-$4.2 & $-$19.8 & $-$24.3 & 35.3 & 9.6 & 55.1 \\
 & PennyLn. & IonQ & $-$19.8 &   0.0 & $-$19.8 &     0.0 & 100.0 & 0.0 & 0.0 \\
\thickhline
\multirow{6}{*}{\textbf{QDRIFT}}
 & TKET      & IBM  & $-$2.2$^\star$ & $-$0.4$^\star$ & $-$2.6 & 0.0 & 84.2$^\star$ & 15.8$^\star$ & 0.0 \\
 & TKET      & IonQ & $-$2.8$^\star$ & $+$0.0$^\star$ & $-$2.8 & 0.0 & 98.6$^\star$ &  1.4$^\star$ & 0.0 \\\cmidrule{2-10}
 & Qiskit    & IBM  & $-$3.4$^\star$ & $-$0.7$^\star$ & $-$4.1 & 0.0 & 81.8$^\star$ & 18.2$^\star$ & 0.0 \\
 & Qiskit    & IonQ & $-$4.4$^\star$ &   0.0$^\star$ & $-$4.4 & 0.0 & 100.0$^\star$ &  0.0$^\star$ & 0.0 \\\cmidrule{2-10}
 & PennyLn. & IBM  & $-$2.6$^\star$ & $-$0.7$^\star$ & $-$3.3 & 0.0 & 79.8$^\star$ & 20.2$^\star$ & 0.0 \\
 & PennyLn. & IonQ & $-$3.3$^\star$ & $+$0.0$^\star$ & $-$3.3 & 0.0 & 99.6$^\star$ &  0.4$^\star$ & 0.0 \\
\hline
\multicolumn{10}{L{15cm}}{
$\star$~H\%/B\% for QDRIFT (and Trotter) are not cross SDK comparable because the H/B boundary is SDK dependent. Use $\Delta F_{H\!+\!B}$ instead. Absolute $F_{\text{tot}}$ (which also includes $T_2$ idle and readout) is outside the physical regime and serves only as a relative compiler overhead proxy.}
\end{tabular}

\end{table}

\begin{figure}[t]
  \centering
  \includegraphics[width=0.75\columnwidth]{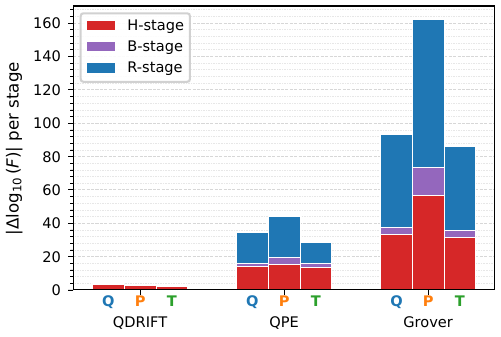}
  \caption{Per-phase fidelity attribution for significant algorithms. R-stage dominates Grover/QPE; QDRIFT is synthesis only ($R$=0).}

  \label{fig:hbr-deep}
\end{figure}

Table~\ref{tbl:hbr-deep} reports the per-stage $\Delta\log_{10}(F_X)$ costs for Grover, QPE and QDRIFT on both IBM (heavy-hex) and IonQ (all-to-all).
Because our objective is phasewise \emph{relative} compiler comparison rather than absolute fidelity prediction, large negative values (e.g., $\log_{10}(F)\ll -1$) should be interpreted as relative overhead proxies, consistent with the model assumptions in \S\ref{sec:model:assumptions}.

\subsubsection{Observations from HBR Attribution}
\label{sec:results:findings}
\paragraph{Routing amplifies synthesis deficits}
On sparse coupling graphs (heavy-hex or a square-lattice), a larger pre-routing circuit produces proportionally more SWAP gates because the gate density itself determines SABRE's layout. Grover on heavy-hex illustrates the mechanism (Tbl.~\ref{tbl:hbr-deep}). PennyLane's \texttt{qml.ctrl(PauliZ)} decomposes MCZ in an ancilla-free recursive manner, producing 38,800 2Q gates after H vs.\ 22,150 for Qiskit (1.75$\times$). After routing on IBM Heron, PennyLane inserts 56,679 SWAPs vs.\ Qiskit's 35,538 (1.60$\times$). QPE shows the same pattern against TKET: PennyLane's SABRE invocation generates 15,525 SWAPs vs.\ TKET's 7,995 (1.94$\times$) from a 1.25$\times$ H+B fidelity deficit (Tbl.~\ref{tbl:hbr-deep}: $-19.8$ vs.\ $-15.9$). The routing ratio therefore does not track the synthesis ratio. It can either compress or amplify the gap depending on circuit structure. We quantify this interaction in \S\ref{sec:discussion}.

\paragraph{The main bottleneck depends on the circuit}
Table~\ref{tbl:hbr-deep} shows two clear categories:
\begin{itemize}[leftmargin=*,nosep]
\item \textbf{Routing-dominated} (R\% $>$ 50\%): Grover and QPE. These circuits use maximally entangled multi qubit gates (e.g., MCX, controlled phase rotations), which are incompatible with sparse hardware topologies. Although IonQ avoids routing due to all-to-all connectivity, the synthesis gap remains (Grover IonQ $\Delta F_{H+B}$: TKET $-39.2$ vs.\ PennyLane $-71.8$, Tbl.~\ref{tbl:hbr-deep}), confirming that H-stage quality is the main issue.
\item \textbf{Synthesis-dominated} (R\% $\approx$ 0\%): QDRIFT and Trotter. These nearest-neighbour TFIM Hamiltonian circuits naturally match the heavy-hex connectivity, so few or no SWAPs are needed. The entire compiler gap resides in H-stage gate reduction.
\end{itemize}

\paragraph{QFT's inter-SDK gap is due to routing.}
All three SDKs reach identical H+B gate counts on QFT ($n$=10, 105~CXeq after basis translation), confirmed on IonQ's routing-free execution where no inter-SDK gap appears ($\log_{10}F$: Qiskit $-0.424$, TKET $-0.423$, PennyLane $-0.439$). On IBM Heron the entire gap is routing-driven: TKET's deterministic \texttt{RoutingPass} inserts 177 SWAPs vs.\ Qiskit's SABRE 141 (1.25$\times$), reflecting router efficiency differences on QFT's dense interaction pattern rather than synthesis quality. The TKET/Qiskit SWAP ratio peaks at $n{=}12$ (321/228 = 1.41$\times$) and converges by $n{\geq}18$ (1.06$\times$ at $n{=}18$, 1.01$\times$ at $n{=}20$).

\paragraph{Five of eight algorithms are compiler-equivalent}
GHZ, QFT, Trotter, BV, and QAOA show fidelity gaps below 0.5d across SDKs (Fig.~\ref{fig:hbr-shallow}). These circuits have low gate counts and simple or nearest-neighbour connectivity, leading to similar compiled results. This validates the HBR methodology in that structurally similar circuits report as equivalent.

\subsubsection{Fidelity Cliff Projections}
\label{sec:results:cliffs}
\begin{figure}[t]
  \centering
  \subfloat[Grover]{\includegraphics[width=0.5\columnwidth]{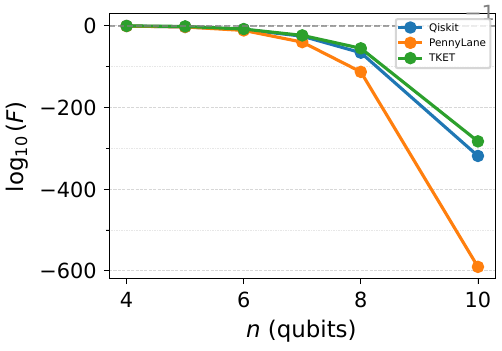}\label{fig:cliff-grover}}%
  \subfloat[QPE]{\includegraphics[width=0.5\columnwidth]{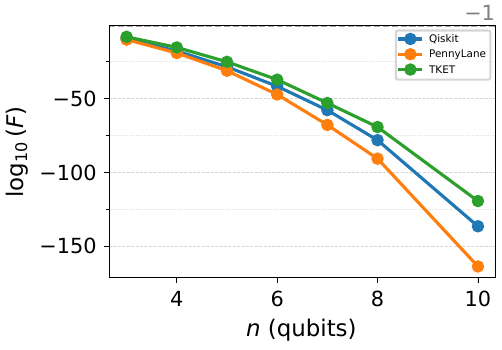}\label{fig:cliff-qpe}}\\
  \subfloat[QFT]{\includegraphics[width=0.5\columnwidth]{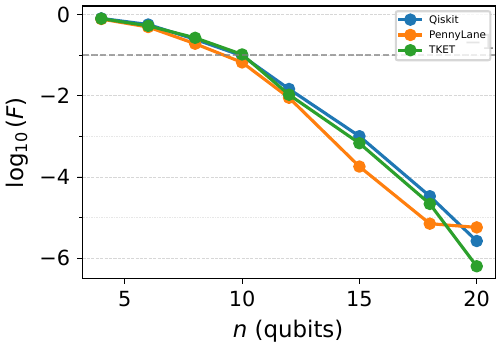}\label{fig:cliff-qft}}%
  \subfloat[QDRIFT]{\includegraphics[width=0.5\columnwidth]{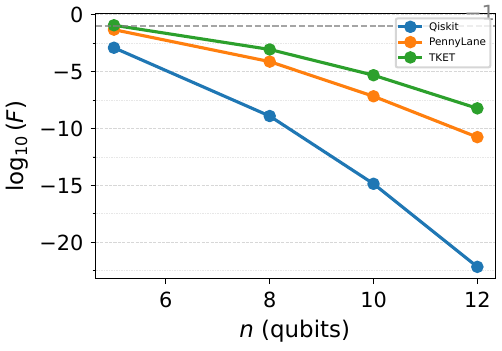}\label{fig:cliff-qdrift}}%
  \caption{$\log_{10}(F)$ vs qubit count ($n$=[3,20]). Dashed line marks the $-1$ threshold.}
  \label{fig:fidelity-cliff}
\end{figure}

Figure~\ref{fig:fidelity-cliff} plots the projected $\log_{10}(F)$ versus qubit count $n\in[3,20]$ under the analytical model, with the dashed line at $-1$ used as a \emph{heuristic} ``fidelity cliff'' threshold for comparing scaling trends across SDKs and backends.

\textbf{Grover} crosses $\log_{10}(F)<-1$ at $n=5$ (all SDKs, both backends, Fig.~\ref{fig:fidelity-cliff}) and is infeasible without QEC~\cite{sergey:nature:2024}.
\textbf{QFT} on IBM crosses the threshold at $n=10$--$12$ ($\log_{10}F=-1.01$ at $n=10$ for Qiskit, $-1.97$ at $n=12$ for TKET). On IonQ it remains viable through $n=15$ ($\log_{10}F=-0.96$) and crosses by $n=18$ ($-1.39$).
\textbf{GHZ, BV} and \textbf{QAOA} remain viable through the tested range $n=20$ ($\log_{10}F$ above $-1$ across all SDKs and backends). \textbf{Trotter} stays viable through the tested $n=12$ ($-0.46$, IBM TKET).
\textbf{QDRIFT} crosses the limit by $n=5$ for Qiskit and PennyLane on IBM ($-2.90$ and $-1.31$ respectively), while TKET remains marginally viable at $n=5$ ($-0.92$) and crosses between $n=5$ and $8$ ($-3.05$). TKET's $\approx $38\% lower 2Q count thus extends the viable range relative to Qiskit at this circuit class.

\subsection{Cross-Tier Validation}
\label{sec:crosstier}

\textit{Tier~1} (\texttt{opt=0}) results provide diagnostic, stage-resolved attribution under strict H/B/R boundaries, whereas \textit{Tier~2} evaluates production-style compilation where passes can interleave across stages.
Because stage boundaries are no longer well-defined at higher optimization levels, Tier~2 is used only for end-to-end comparisons, not for phasewise attribution.

\subsubsection{Setup}
\label{sec:crosstier:setup}
At \texttt{opt=0}, HBR attribution enforces strict stage boundaries; at opt$\ge 2$, production pipelines may rewrite gates after basis translation and/or after routing, mixing optimization and translation so that a phasewise allocation is no longer meaningful.
For production relevance, we compare three pipelines that are representative of how users compile circuits in practice: (i) \textbf{Qiskit-opt2:} \texttt{transpile(optimization\_level=2)}, a multi-pass production pipeline including routing and post-routing optimizations;
(ii) \textbf{PennyLane-opt2:} PennyLane front-end compilation (H/B as in Tier~1) followed by Qiskit \texttt{transpile(optimization\_level=2)} for routing and post-routing cleanup; and
(iii) \textbf{TKET-default:} TKET's default full compilation pipeline.
PennyLane-opt2 uses the same circuit construction as Tier~1 (e.g., \texttt{qml.ctrl(qml.PauliZ)} in Grover), so any front-end synthesis differences persist into Tier~2, while routing and post-routing optimization are handled by Qiskit in a comparable manner across Qiskit-opt2 and PennyLane-opt2.

\subsubsection{Observations from Crosstier results}
\label{sec:crosstier:findings}

\begin{table}[t]
\centering
\caption{Crosstier consistency, rank-1 (gap) on IBM at $n$=10}
\label{tbl:cross-tier}
\begin{tabular}{l|lll}
\hline
\bf Algo. & \bf opt=0  & \bf opt=2 & \bf Result \\
\hline\hline
\bf GHZ & $\approx$All (0.0d) & $\approx$All (0.0d) & Consistent \\
\bf Grover & TKET (35.6d) & \textbf{QK} (63.3d) & \textbf{Reversal} \\
\bf QDRIFT & TKET (1.9d) & $\approx$QK (0.0d) & \textbf{Converge} \\
\bf QFT & $\approx$All (0.0d) & $\approx$All (0.1d) & Consistent \\
\bf QPE & TKET (17.1d) & \textbf{PL} (9.0d) & \textbf{Reversal} \\
\bf Trotter & $\approx$All (0.0d) & $\approx$All (0.0d) & Consistent \\
\bf BV & $\approx$All (0.0d) & $\approx$All (0.0d) & Consistent \\
\bf QAOA & $\approx$All (0.0d) & $\approx$All (0.0d) & Consistent \\
\hline
\end{tabular}
\end{table}

\begin{figure}[t]
  \centering
  \includegraphics[width=0.75\columnwidth]{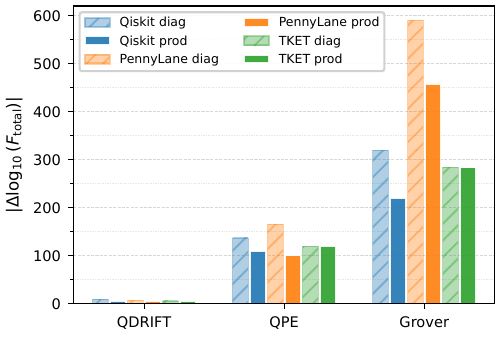}
  \caption{Optimization level winner, comparing Tier 1 (diag) vs Tier 2 (prod). Grover/QPE reverse; QDRIFT converges.}

  \label{fig:cross-tier}
\end{figure}

Table~\ref{tbl:cross-tier} and Fig.~\ref{fig:cross-tier} show three recurring patterns.

\paragraph{Grover}
At \texttt{opt=0}, TKET ranks first, consistent with Tier~1 routing dominance. At \texttt{opt=2}, Qiskit ranks first by 63.3\,d. Qiskit's
multipass pipeline reduces Grover to 43,975 2Q gates versus TKET's 54,273. PennyLane's MCZ synthesis deficit persists at \texttt{opt=2} (83,477 2Q, $1.9\times$ Qiskit), an H-stage gap that post-routing optimization cannot close. This reversal shows that \texttt{opt=0} leaders do not transfer to production.

\paragraph{QPE}
TKET wins at \texttt{opt=0}, reversing to PennyLane at \texttt{opt=2} by 9.0\,d. PennyLane-opt2 produces marginally more 2Q gates than
Qiskit-opt2 (9,800 vs.\ 9,665) yet achieves better fidelity ($-112.2$ vs.\ $-117.8$\,d). The cause is depth: PennyLane-opt2's shallower circuit (10,844 vs.\ 12,628 layers) reduces idle qubit time and the $T_2$ penalty, consistent with Eq.~\ref{eq:total-fidelity}. Two pipelines with similar 2Q counts can therefore differ in total fidelity when depth diverges.

\paragraph{QDRIFT}
All three SDKs converge to within 0.3\,d at \texttt{opt=2}. Qiskit's multipass pipeline replicates the gate reduction TKET achieves by default, erasing the Tier~1 gap. Production-level optimization can therefore erase synthesis differences that are visible under strict diagnostic boundaries.

\paragraph{Remaining circuits}
GHZ, QFT, Trotter, BV, and QAOA remain equivalent at both tiers ($<$0.1\,d gap, Table~\ref{tbl:cross-tier}). These circuits are either shallow or structurally constrained such that production optimizers find no inter-SDK headroom.

\begin{figure}[t]
  \centering
  \includegraphics[width=0.75\columnwidth]{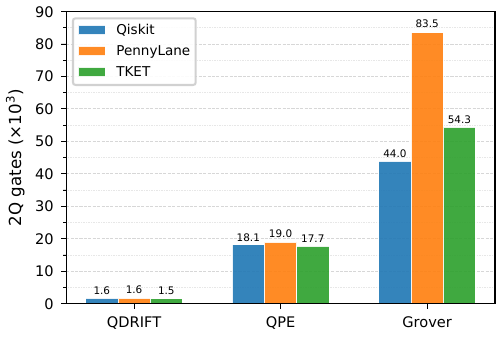}
  \caption{Total 2Q gate counts by SDK for production (Tier 2) compilation (\texttt{opt=2}/default).}

  \label{fig:production-2q}
\end{figure}

\subsubsection{Compile Time Overhead}
\label{sec:crosstier:compile}

\begin{figure}[t]
  \centering
  \includegraphics[width=0.75\columnwidth]{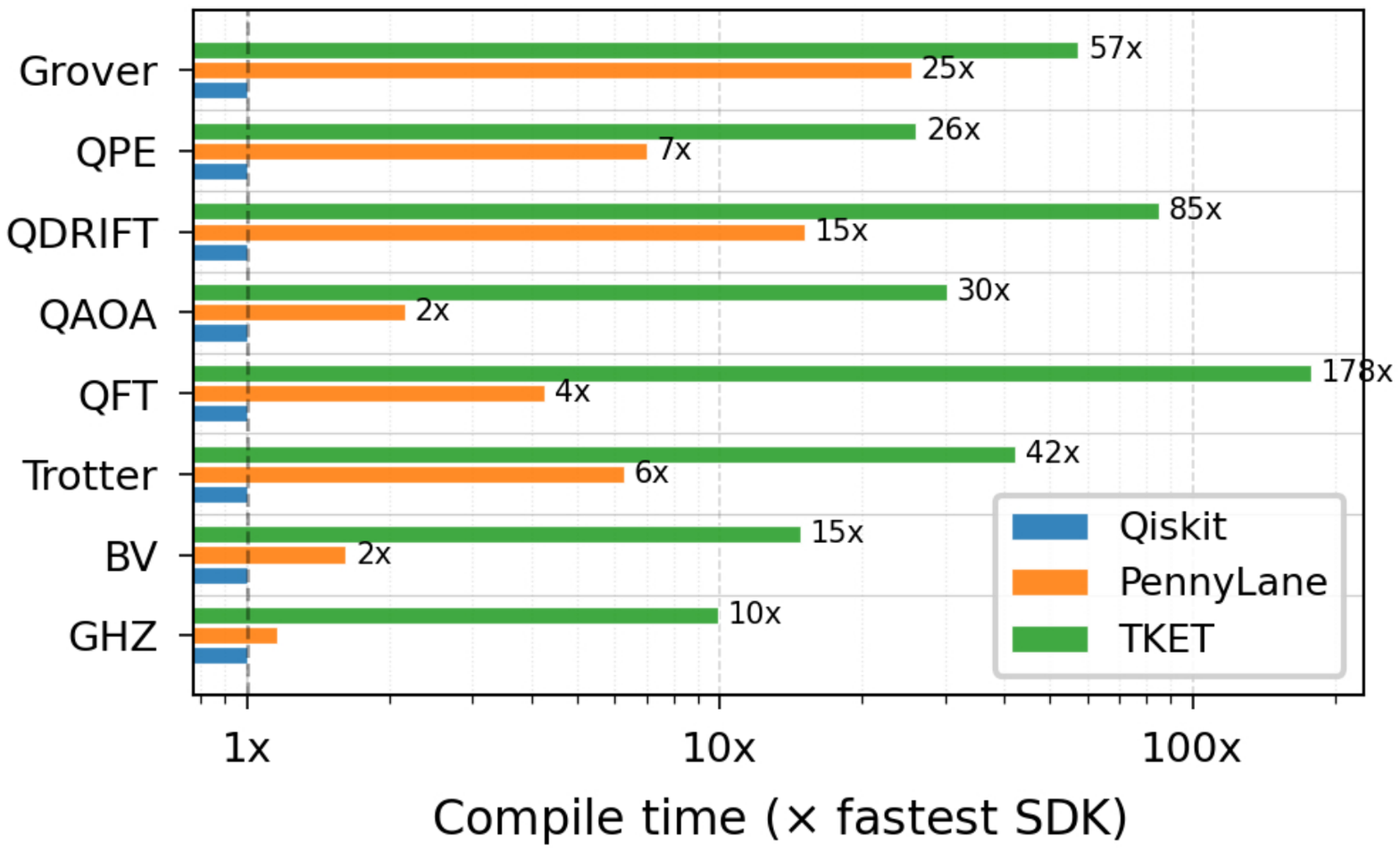}
  \caption{Tier~2 compile time (ms) by SDK}

  \label{fig:compile-time}
\end{figure}

Production compilation introduces a classical runtime tradeoff: more aggressive pipelines can reduce quantum circuit cost but incur higher compile-time overhead.
TKET's overhead ranges from 10$\times$ (GHZ) to 57$\times$ (Grover) relative to Qiskit-opt2, indicating that this is highly dependent on circuits and is worst for circuits with repeated rotation blocks vulnerable to deep peephole analysis. Figure~\ref{fig:compile-time} shows production level compile time overhead with Qiskit as the baseline. This should be considered alongside fidelity improvements when selecting a production pipeline.

\subsection{Model Validation}
\label{sec:validation}

We validate if the analytical model preserves \emph{SDK rank ordering} under realistic noisy simulators and real hardware executions.
Since the model is intended primarily for relative comparisons, the validation criterion is \textit{rank agreement} between the predicted fidelity proxy $F_{\text{pred}}$ and measured success metrics under consistent experimental conditions.
Unless noted otherwise, ties are handled using the same detection rule as Table~\ref{tbl:sim-validation} ($\varepsilon=0.01$ predicted; $2\sigma$ measured).

\subsubsection{IBM FakeFez Noisy Simulator}
\label{sec:validation:sim}

All 8 benchmark circuits are executed across all SDK/tier variants on \texttt{AerSimulator} with the \texttt{FakeFez} noise snapshot, which
reproduces per-edge CZ, per-qubit SX, readout, $T_1$, and $T_2$ calibration data for \texttt{ibm\_fez} (Heron r2, 156 qubits, heavy-hex). Each variant is compiled to the native \{CZ, SX, X, RZ\} basis and routed to the FakeFez coupling map (SABRE for Qiskit and PennyLane, \texttt{GraphPlacement} for TKET) and executed at 8{,}192 shots. Edges with CZ error above $0.10$ are removed from the coupling map before transpilation; such edges correspond to anomalous couplers with error ${\approx}0.80$ that would otherwise attract connectivity-aware routers (TKET's \texttt{GraphPlacement}) and produce deterministically poor circuits.

For circuits with a single correct output, $P_{\text{ibm}}$ is the fraction of shots on the correct bitstring. For distributional outputs
(QFT, Trotter, QDRIFT), we use Hellinger fidelity. The predictor $F_{\text{pred}}$ uses the validation configuration (\S\ref{sec:model:variants}) with per-edge parameters extracted from Aer's noise channels via \texttt{average\_gate\_fidelity}.

\noindent\textbf{Outcome.} All 8 circuits achieve rank agreement between $F_{\text{pred}}$ and $P_{\text{ibm}}$ across both tiers (Table~\ref{tbl:sim-validation}, IBM columns). Across the 6 SDK/tier variants per circuit, the mean within-circuit Pearson correlation is $r{=}0.993$.

\begin{table}[t]
\centering
\caption{Noisy-simulator validation across two platforms and 2 optimization tiers: IBM FakeFez (heavy-hex, 8{,}192 shots) and IonQ Forte-1 (all-to-all, 1{,}024--8{,}192 shots). $N_{2Q}$ range spans all 6 SDK/tier variants. Rank match uses tie detection ($\varepsilon{=}0.01$ predicted, $2\sigma$ measured).}
\label{tbl:sim-validation}
\def\thickhline{\noalign{\hrule height1pt}}
\begin{tabular}{l|r|rr|c|rr|c}
\hline
\bf Circuit & \bf $n$ & \multicolumn{3}{c|}{\bf IBM FakeFez} & \multicolumn{3}{c}{\bf IonQ Forte-1} \\ \cline{3-8}
 & & $N_{2Q}$ & $P_{\text{ibm}}$ & \em Rank & $N_{2Q}$ & $P_{\text{ionq}}$ & \em Rank \\
\hline\hline
\multicolumn{8}{c}{\cellcolor{gray!20}\emph{Synthesis-discriminating}} \\\thickhline
Grover  &  4 & 111--264 & .20--.51 & \checkmark & 84--120 & .30--.47 & \checkmark \\
QFT     &  4 & 21--54   & .73--.89 & \checkmark & 12--36  & 1.0      & \checkmark$^\dagger$ \\
QPE     & 3+1 & 53--74  & .98--1.0 & \checkmark$^\dagger$ & 35--46  & 1.0 & \checkmark$^\dagger$ \\
QDrift  &  4 & 16--20   & .96--1.0 & \checkmark & 16--20  & .97--1.0 & \checkmark \\
\hline
\multicolumn{8}{c}{\cellcolor{gray!20}\emph{SDK-equivalent on at least one platform}} \\\thickhline
BV      & 12 & 13--21   & .71--.86 & \checkmark & 6 (all) & 1.0 & \checkmark$^\dagger$ \\
QAOA    &  8 & 44--57   & .95--.97 & \checkmark & 24 (all) & .83--.86 & \checkmark \\
GHZ     &  8 & 7 (all)  & .79--.89 & \checkmark & 7 (all)  & .93--.96 & \checkmark \\
Trotter &  4 & 24 (all) & .37--.56 & \checkmark & 24 (all) & .52--.59 & \checkmark \\
\hline
\multicolumn{8}{l}{$^\dagger$\,Near-saturation ($P{\geq}.98$). Rank match via ties.}
\end{tabular}
\end{table}

\subsubsection{IonQ Forte-1 Noisy Simulator}
\label{sec:validation:forte1}
We validate the model on a second hardware class by executing the same set of compiled circuits on IonQ's cloud Forte-1 noisy simulator (\texttt{ionq\_simulator}, \texttt{noise\_model="forte-1"}), which provides an all-to-all topology representative of trapped-ion connectivity. All-to-all connectivity sets $\Delta F_R{=}0$ by construction, so inter-SDK differences reduce to pre-routing compilation choices captured by
the H and B stages (Eq.~\ref{eq:total-fidelity}).

The attribution model uses Markovian depolarizing noise (Sec.~\ref{sec:model}). Depolarizing models built from component benchmarks are known to overpredict absolute fidelity on Forte-class systems~\cite{chen:quantum:2024}, and the Forte-1 simulator incorporates additional non-idealities beyond this abstraction. Absolute $P_{\text{ionq}}$ can therefore fall below $F_{\text{pred}}$ for deep circuits. Rank preservation nonetheless holds: on an all-to-all topology every SDK executes on the same physical qubits, so any noise penalty that scales monotonically
with circuit cost preserves ordering.

\textit{All eight circuits achieve rank agreement} between $F_{\text{pred}}$ and $P_{\text{ionq}}$ (Table~\ref{tbl:sim-validation}, right columns) under the same tie rule as IBM validation. Grover is the primary discriminator: Qiskit and TKET compile to $N_{2Q}{=}84$ while PennyLane produces $N_{2Q}{=}120$ via its \texttt{qml.ctrl(qml.PauliZ)} oracle construction. QFT, QPE, and BV saturate at $P_{\text{ionq}}{\geq}0.98$ and are tied within shot noise. QAOA, GHZ, and Trotter produce identical $N_{2Q}$ across all three SDKs on all-to-all connectivity and match as expected.

\subsubsection{Hardware Validation on IBM Fez}
\label{sec:validation:hardware}

The model is validated on \textbf{IBM Fez} (Heron r2, 156 qubits) using two circuits. \textbf{BV n=12} (13 qubits, secret=\texttt{110010100110}, 8,192 shots, Job~\texttt{d6v42vitnsts73esq360}) and \textbf{Grover n=4} (4 qubits, marked $|1111\rangle$, $k$=3, 8,192 shots, Job~\texttt{d7gsdeq2khts739ounig}). BV provides a deterministic correctness check whereas Grover adds a deeper circuit. The validating metric is whether the predicted SDK ordering matches the measured $P_{\text{correct}}$.

\textit{The predicted ordering matches the measured ordering for both circuits at both tiers} (Table~\ref{tbl:hw-validation}).

\textbf{BV.} At Tier~2, Qiskit-opt2 and PennyLane-opt2 produce identical circuits ($N_{2Q}{=}15$, $D_{\text{tot}}{=}26$). TKET matches on $N_{2Q}$ but incurs larger depth ($32$), resulting in a ${\approx}5$\,pp lower $P_{\text{hw}}$ that falls within binomial shot-noise uncertainty ($\sigma \approx 0.54\%$ at $8{,}192$ shots). The dominant penalty is idle $T_2$ decoherence from depth (Eq.~\ref{eq:decoherence}), not raw 2Q count, directly validating the inclusion of both terms in Eq.~\ref{eq:total-fidelity}.

\textbf{Grover.} Qiskit ranks first at both tiers via the most compact MCX decomposition ($N_{2Q}{=}135$ at Tier~1, $111$ at
Tier~2). At Tier~1, TKET produces far fewer 2Q gates than PennyLane ($138$ vs.\ $243$), yet achieves lower hardware fidelity. The cause
is qubit placement: TKET's \texttt{GraphPlacement} routes through physical qubits $\{98, 110, 111, 112\}$, where q98 has
$T_2{=}18.6\,\mu$s against $82$--$151\,\mu$s for the remaining mapped qubits, consistent with TLS-dominated dephasing. The per-qubit
model attributes $73\%$ of TKET's total $T_2$ penalty to q98. The residual gap between TKET and PennyLane is $0.006$ (${\approx}1.2\sigma$), within shot-noise uncertainty, indicating the Markovian model slightly underpredicts the penalty on the degraded qubit. Hardware-aware placement and qubit-level QEC are outside the scope of this work.

\begin{table}[t]
\centering
\def\thickhline{\noalign{\hrule height1pt}}

\caption{Hardware validation on IBM Fez. $F_{\text{pred}}$ = predicted fidelity; $P_{\text{hw}}$ = measured success probability. Rank ordering matches for both circuits at both tiers.}
\label{tbl:hw-validation}
\begin{tabular}{l|r|l|r|r|r|r}
\hline
\bf Circuit & \bf Rank & \bf SDK & $N_{2Q}$ & \bf $D_{\text{tot}}$ & \bf $F_{\text{pred}}$ & \bf $P_{\text{hw}}$ \\
\hline\hline
\multicolumn{7}{c}{\cellcolor{gray!20}\emph{Tier 1 \texttt{(opt=0)}}} \\
\thickhline
\bf BV $n=12$     & 1 & TKET    & 15  & 32  & .644 & .607 \\
              & 2 & Qiskit  & 18  & 42  & .592 & .601 \\
              & 3 & PL      & 18  & 42  & .592 & .596 \\
\cmidrule{1-7}
\bf Grover $n$=4  & 1 & Qiskit  & 135 & 519 & .369 & .409 \\
              & 2 & TKET    & 138 & 360 & .268 & .186$^\dagger$ \\
              & 3 & PL      & 243 & 712 & .227 & .192$^\dagger$ \\
\hline
\multicolumn{7}{c}{\cellcolor{gray!20}\emph{Tier 2 (production)}} \\
\thickhline
\bf BV $n=12$     & 1 & QK-opt2 & 15  & 26  & .664 & .654 \\
              & 2 & PL-opt2 & 15  & 26  & .664 & .644 \\
              & 3 & TKET    & 15  & 32  & .644 & .605 \\
\cmidrule{1-7}
\bf Grover $n=4$  & 1 & QK-opt2 & 111 & 300 & .499 & .473 \\
              & 2 & PL-opt2 & 224 & 523 & .298 & .328 \\
              & 3 & TKET    & 138 & 360 & .268 & .193 \\
\hline
\end{tabular}
\end{table}

\subsubsection{Noise Model Sensitivity}
\label{sec:validation:sensitivity}
To assess whether SDK rankings depend sensitively on calibration accuracy, we perturb the hardware model by increasing the error rates of the worst $10\%$ coupling edges in FakeFez by $20\%$ (16 edges).
Routing-dominated circuits exhibit the largest absolute degradation (e.g., Grover: $\Delta\log_{10}F=-19.45$, QPE: $-0.64$), whereas synthesis-dominated TFIM workloads (QDRIFT, Trotter-Suzuki) remain nearly unchanged ($|\Delta| \le 0.02$).
This confirms that the model responds strongly to perturbations on difficult routing paths rather than to uniform gate-count changes and that the \emph{relative} SDK ordering remains stable under such calibration variations.

\subsubsection{Statistical Validity}
\label{sec:statistical}
Among the three HBR stages, only routing (SABRE) is stochastic; synthesis (H) and basis translation (B) are deterministic under our instrumentation.
TKET's \texttt{GraphPlacement} and \texttt{RoutingPass} stage is also deterministic ($\sigma=0$ across seeds).
To quantify routing-induced variance, we recompute all SDKs across 25 transpiler seeds (16, 18, 20, \ldots, 64) with a 300\,s timeout per seed.

In Table~\ref{tbl:seed-variance}, algorithms marked $*$ are fully deterministic, with gaps below the comparability threshold ($\approx 0.5$ decades).
For the other benchmarks, a winner is defined only when the inter-SDK gap exceeds both the $0.5$-decade comparability threshold and $2\sigma$ of routing noise.
Under this criterion, \textit{all statistically evaluated winners are robust to SABRE's stochasticity.}

\begin{table}[t]
\centering
\caption{25-seed transpiler variance on IBM ($n{=}10$, \texttt{opt=0}).}
\label{tbl:seed-variance}
\begin{tabular}{l|crrc}
\hline
\bf Algo. & \bf Winner &\bf  Gap (d) & \bf Gap/$\sigma$ & \bf Sig? \\
\hline\hline
\bf GHZ & $\approx$All & 0.000 & $\infty$ & $*$ \\
\bf Grover & TKET  & 10.72 & $\gg 1$ & Yes \\
\bf QDRIFT & TKET & 0.80 & $\infty$ & Yes \\
\bf QFT & $\approx$QK & 0.008 & 3.2 & Yes \\
\bf QPE & TKET  & 1.48 & 16.1 & Yes \\
\bf Trotter & $\approx$QK & 0.003 & $\infty$ & $*$ \\
\bf BV & $\approx$TKET & 0.002 & 6.0 & Yes \\
\bf QAOA & $\approx$QK & 0.006 & 11.5 & Yes \\
\hline
\end{tabular}
\end{table}

\section{Discussion}
\label{sec:discussion}

HBR decomposition exposes when routing preserves vs.\ amplifies pre-routing synthesis/translation gaps.
We quantify this interaction for an SDK pair $(a,b)$ (with $a$ the weaker synthesiser) using the routing amplification factor
\begin{equation*}
\label{eq:routing-amplification}
\rho^{(a \to b)} =
\frac{\Delta\log_{10}(F_{H\!+\!B\!+\!R})^{(a \to b)}}
     {\Delta\log_{10}(F_{H\!+\!B})^{(a \to b)}},
\qquad
\Delta X^{(a\to b)} = |X_a - X_b|.
\end{equation*}
$\rho=1$ indicates routing preserves the pre-routing gap, while $\rho>1$ indicates amplification. Table~\ref{tbl:hbr-deep} gives $\rho=1.92$ (Grover), $\rho=4.01$ (QPE) and $\rho=1.00$ (QDRIFT), matching the circuit-class dependence in \S\ref{sec:results:findings}.

\subsubsection{Implications for compiler design}
First, H-stage aggressiveness should be adaptive: the same optimization depth can help structured Hamiltonian kernels but add overhead elsewhere, suggesting feature-conditioned synthesis policies rather than fixed pipelines~\cite{li:asplos:2022,quetschlich:tqc:2025}.
Second, production pipelines can erase diagnostic gaps at substantial compile-time cost, motivating targeted, low-cost passes for common simulation motifs (e.g., ZZ+X blocks) and tighter integration of domain-aware techniques from simulation compilers~\cite{li:asplos:2022,paykin:arxiv:2023,yang:arxiv:2025}.

\subsubsection{Scope and limitations}
Tier~1 (\texttt{opt=0}) results are diagnostic. Optimization-level semantics differ across SDKs, so Tier~1 winners are not production
recommendations. Hardware runs include placement effects because routers may select different physical qubits. CXeq (Table~\ref{tbl:cx-equiv})
assumes maximally entangling native 2Q gates. For fractional-entangling primitives or heterogeneous ISAs (e.g.\ $\sqrt{\text{iSWAP}}$~\cite{%
arute:nature:2019}, ISA-aware routing~\cite{canopus:arxiv:2025}), the weights require revision.

\section{Conclusion}
\label{sec:conclusion}

In this paper, we presented \emph{HBR decomposition}, a per-phase fidelity attribution methodology for quantum compiler pipelines, and evaluated it across 3 SDKs $\times$ 8 algorithms $\times$ 2 backend topologies.
We find that (i) routing dominates the compilation-induced fidelity cost for search-class circuits on sparse connectivity (e.g., Grover, QPE), whereas Hamiltonian simulation workloads are primarily synthesis/translation limited (e.g., QDRIFT, Trotter); (ii) early H-stage choices can amplify or compress downstream routing overhead depending on circuit structure; and (iii) diagnostic winners at \texttt{opt=0} can reverse under production compilation (\texttt{opt=2}/default), confirming that stagewise attribution and production performance answer different questions.
The model preserves SDK rank ordering under validation on IBM FakeFez and IonQ Forte-1 noisy simulators and on IBM Fez hardware executions, supporting its use as an actionable diagnostic for identifying the dominant compilation bottleneck for a given circuit family and topology.
Future work includes extending HBR to additional SDKs and application domains (e.g., quantum chemistry), incorporating dynamical decoupling into the $T_2$ term, and studying interactions with error mitigation.

\clearpage

\bibliographystyle{plain}

\bibliography{paper}

\end{document}